\documentclass[%
 reprint,
 amsmath,amssymb,
 aps,
]{revtex4-2}

\usepackage{graphicx}
\usepackage{dcolumn}
\usepackage{bm}

\begin{document}

\title{Geometry aware predictive models for exocytosis}

\author{Sundeep Kapila}
 \altaffiliation {Electrical Engineering Department, IIT Bombay}\email{214074001@iitb.ac.in}
\author{Pradeep R. Nair}%
\altaffiliation {Electrical Engineering Department, IIT Bombay}
 \email{prnair@ee.iitb.ac.in}
\affiliation{%
Department of Electrical Engineering, Indian Institute of Technology Bombay, Mumbai, India
}%

\begin{abstract}
Inter neuron communication happens through the exchange of neurotransmitters at the synapse by a process known as exocytosis. This makes exocytosis a fundamental process of information exchange in the body. The exocytosis process has a distinct geometry as it involves a vesicle that attaches to the cell membrane and then releases the neurotransmitters through a pore. Significant recent research, both experimental and numerical, attempt to understand the time dynamics of exocytosis. In this manuscript, we share an analytical model that predicts the key output parameters of exocytosis based on the geometry of the vesicle and pore. Our analytical predictions are well supported by detailed numerical simulations. This model could help extract geometrical parameters from experimental data and hence could be of broad interest.

\end{abstract}

\maketitle

\section{\label{sec:intro}Introduction}
It is well known that information exchange between neurons is mediated through transfer of neurotransmitters\cite{intro1}. Accordingly, the process of exocytosis in which vesicles release neurotransmitters to the pre-synaptic cleft is of fundamental importance. Extensive research, both experimental and numerical \cite{analytical1} \cite{analytical2}, has been reported to analyze the time dynamics of exocytosis and hence decipher the underlying physical processes (like pore opening, closure, kiss and run, etc.). Key experimental approaches involve electrochemical detection of released neurotransmitters \cite{nano1} \cite{mos} \cite{biovesicles} and optical methods like super-resolution stimulated emission depletion (STED) microscopy \cite{sted}, while numerical approaches are routinely used to gather insights on the time dynamics of exocytosis. Accordingly, pore dynamics is understood to have a critical role on the transient exocytosis and associated time constants \cite{pore_exp}. For example, early literature reports anticipated that the time constants to be near universal and independent of pore width. Later reports attribute pore closure phenomena to the emergence of bi-exponential trends in time dynamics. Interestingly, recent experiments indicate that a significant fraction of exocytosis events do not involve pore closure \cite{sted}.\\

Despite the above mentioned advances, several key features related to the time dynamics of exocytosis event are yet to be quantitatively established. For example, (i) How are the time constants related to the size of vesicles and pore? (ii) How does pore dynamics (especially the closure event) influence the time constants?, and  (iii) Is it possible to delineate the physical events through a critical analysis of the transient data? To this end, here we use a combination of predictive analytical models and detailed numerical simulations to explore the influence of various physical events that contribute to exocytosis. Our results clearly highlight several interesting trends. For example, we find that the physical dimensions of the pore cause a transition from single exponential to multi-exponential decays. We identify functional dependence of key features like the peak time, peak flux, quantal and time constants on the geometry/shape of vesicles and pores. This allows us to propose new analysis schemes for direct and facile back extraction of information regarding the geometry/shape of vesicles and pores. \\

Below, we first describe the model system to explore the time dynamics of excocytosis. Key features of the time dynamics are then identified through analytical modeling. Through detailed numerical simulations, we then validate the analytical predictions which allows us to identify schemes/methods to back extract the key features like vesicle and pore size, etc. 

\section{Model System}

\begin{figure}[ht]
\includegraphics[width=\columnwidth]{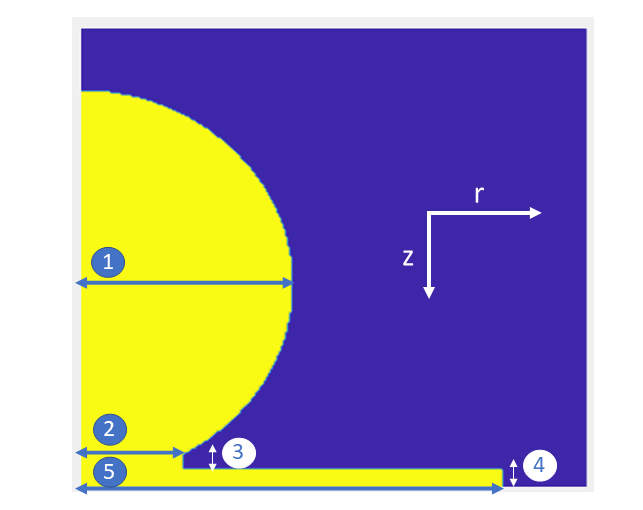}
\caption{\label{fig:model_cylinder} The model layout and key parameters (1) Vesicle Radius (typically 100 nm) (2) Pore Width (15 to 60 nm) (3) Pore Height (7-10 nm) (4) Sensor Distance (7 to 15 nm) (5) Sensor Length (200 nm)}
\end{figure}

In exocytosis, neurons transfer special bio-molecules called neurotransmitters \cite{intro1}.  Neurotransmitter  molecules are grouped together in a vesicle and travel through the axon. On reaching the synaptic terminal of the sender neuron, the vesicles fuse with the axon membrane and release the neurotransmitters in the synaptic cleft through a pore at the vesicle and membrane interface. These neurotransmitters travel in the cleft to the receptors on the dendrite's membrane of the receiver neuron.The rate of transfer of neurotransmitters in the synaptic cleft has been recorded by placing carbon nano electrodes in the synaptic cleft \cite{nano1}. The typical profile of the signal recorded is a sharp rise followed by a steady fall. \\
The time dynamics of exocytosis could be contributed by several phenomena like pore opening, diffusion of neurotransmitters, pore closing, etc. However, a recent study indicated that the pore width remains unchanged in a significant fraction of excocytosis events \cite{sted}. Accordingly, here we explore the time dynamics of an exocytosis event where the pore width remains unchanged. As shown in Fig. \ref{fig:model_cylinder}, our model system consist of a spherical vesicle attached to a membrane with a cylindrical pore and a planar sensor at a distance $d$ from membrane.  We assume the vesicle is uniformly filled with a concentration $C_s$ and at t=0, the pore is opened and the neurotransmitters are allowed to diffuse out, first through the cylindrical pore height and then in the extracellular medium towards the sensor.  Accordingly, the key parameters which define the geometry and hence influences the time dynamics of exocytosis events are geometric parameters like vesicle radius $(r)$, pore width $(w)$, pore height $(h)$ and sensor distance $(d)$, see Fig. ~\ref{fig:model_cylinder}. 

Employing cylindrical symmetry of the system, diffusion of the neurotransmitter from the vesicle to the cleft and further detection at the electrode is modeled using the diffusion equation as follows

\begin{equation} \label{eq:diff_eq2}
\frac{\delta C}{\delta t} = D \left[ \frac{1}{r}\frac{\delta C}{\delta r} +\frac{\delta^2 C}{\delta r^2} + \frac{\delta ^2 C}{\delta z^2} \right]
\end{equation}
where D is the diffusion constant (assumed constant). The variables $r$ and $z$ are as shown in Fig. ~\ref{fig:model_cylinder}.\\

\begin{figure}[ht]
\includegraphics[width=\columnwidth]{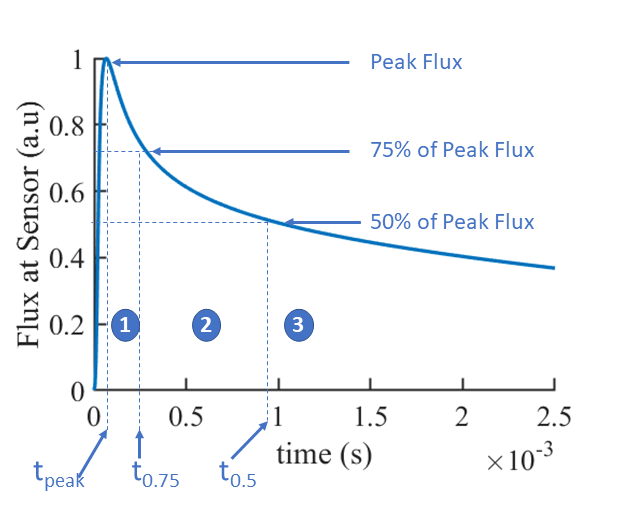}
\caption{\label{fig:output_parameters} Typical output signal and output parameters analysed}
\end{figure}

The typical signal measured at the sensor post the diffusion is as shown in Fig. ~\ref{fig:output_parameters}. As the pore opens at $t=0$, neurotransmitters diffuse through the pore and reach the sensor. Electrochemical reaction at the sensor results in a current which is proportional to the net flux of neurotransmitters at the sensor surface. Evidently, the current increases sharply as the pore opens due to the sudden  influx of neurotransmitters. With time the concentration of molecules in the extracellular medium increases and this reduces the concentration gradient across the pore and hence the diffusive flux of neurotransmitters from the vesicle decreases. As a result, the sensor current exhibits a decreasing trend after reaching a peak.  Accordingly, the key parameters that  characterise the important features of such measured signals are:  (i) Peak flux ($F_{peak}$) as a measure of the maximum magnitude of the signal and defined as the peak of the signal, (ii) Peak Time ($t_{peak}$) as the time taken to reach the peak flux, (iii) Time constants ($\tau$) to characterise the decay of the signal post attaining peak, and (iv) Quantal to characterise the total number of neurotransmitters captured by the sensor.

\section{Predictive models for output parameters}
Based on equation \ref{eq:diff_eq2}, here we develop an analytical approach that predicts the functional dependence of peak flux, peak time and the time constants on the geometry of vesicle and pore. Later these predictions are compared against detailed numerical simulations.

\textbf{A. Time constants} During exocytosis, neurotransmitters are released into the synaptic cleft which are later sensed by the electrode. As described earlier, the sensor current shows a decreasing trend after a sudden initial increase. In a typical scenario, we identify  three different regimes in the transients post peak and consequently three different time constants could be of interest. The different regimes are: (a)  the decay in signal from peak to reaching 75\% of peak (with a time constant $\tau_{head}$), (b)  decay in signal from 75\% of peak to
50\% of peak (with a time constant $\tau_{mid}$), and (c)  decay
from 50\% of peak onwards (with a time constant $\tau_{mid}$). These regions are depicted in Fig. ~\ref{fig:output_parameters}. Quick estimates for the time constants can be obtained through careful analysis as follows.\\

The first regime is when there is build up of concentration in the pore. In this regime, the flux at sensor reaches a peak and then as the build up in the pore is about to reach a steady state, the flux at sensor starts dropping. This we have assumed till the flux reaches $75\%$ of the peak flux. This short period of decay post reaching the peak flux is characterised by a time constant labeled as $\tau_{head}$. 

In the second regime, the concentration in the pore has reached a steady state, but the concentration in the extracellular fluid between the pore and the sensor has not reached steady state.  There are two factors acting, (i) the particles are reaching more part of the sensor through radial diffusion. (ii) reduction  of the flux at the pore opening. The first factor would increase the total flux at the sensor, while the second reduces the absolute flux at each point on the sensor. The net result is that the flux at the sensor decays with time with a time constant which is largely based on pore height as that is the bottleneck in the diffusion process.  We have assumed this regime to last till the flux reaches $50\%$ of the peak flux and characterised it with the time constant labeled as $\tau_{mid}$.

The third and final regime is when the diffusion has reached a steady state both in the pore height and in the extracellular space between the pore and the sensor. In this regime, both the pore height and the gap between the pore and sensor become a bottleneck and the diffusion is limited by both of these parameters. The decay is characterised with a time constant labeled as $\tau_{tail}$.

The derivation and expression for the time constants in the $mid$ and $tail$ regimes are similar as both of them are fundamentally the same diffusion process, although limited by different parameters due to the geometry of the system. The $mid$ regime can be seen as a 1D diffusion along a cylinder and is limited by the pore height $(h)$. The $tail$ regime can be seen as a combination of 1D diffusion along a cylinder followed by a 2D diffusion in space, and is limited by both the pore height $(h)$ and the sensor distance $(d)$. Hence, to derive an analytical expression for the time constants in the $mid$ and $tail$ regime, we follow the same steps and in the end apply the limiting condition to get the final expressions. 

We can make the following assumptions (i) the flux at the sensor is directly proportional to the flux at the pore opening, (ii) the flux at the pore opening has reached a steady state which can be approximated by a linear concentration profile across the cylindrical pore.
Based on the above, we can derive an equation for the rate of change of concentration at the pore opening at a given time as follows, where $D$ is the diffusion constant, $w$ is the pore width, $h$ is the pore height, $r$ is the vesicle radius, $C(z,t)$ denotes concentration at time $t$, at co-ordinate $z$ along the $z$ direction with $z=0$ at the pore and vesicle interface and $z=h$ at the pore and external interface:
\begin{equation} \label{eq:flux_time}
    Flux(h,t) \approx \pi w^2 D \frac{\left[ C(h,t) - C(0,t) \right]}{h}
\end{equation}

Based on conservation of total number of particles, we can write
\begin{equation}
    C(h,t) = C(0,0) - \frac{3}{4 \pi r^3} \int_{0}^t Flux(h,t)dt
\end{equation}
Differentiating the above equation, we get
\begin{equation} \label{eq:conservation}
    \frac{\delta C(h,t)}{\delta t} = \frac{3}{4 \pi r^3}Flux(h,t)
\end{equation}
Substituting the expression for $Flux(h,t)$ from equation \ref{eq:flux_time} in equation \ref{eq:conservation}, we get
\begin{equation} \label{eq:tau_eqn}
    \frac{\delta C(h,t)}{\delta t} = \frac{3 w^2 D}{4r^3 h} \left[ C(h,t) - C(0,t) \right]
\end{equation}
The above is a linear differential equation with a time constant given by
\begin{equation} \label{eq:tau_final}
    \tau = \frac{4r^3 h}{3 w^2 D}
\end{equation}

Now, in the $mid$ regime, the limiting factor is the pore height $(h)$, and hence the expression for $\tau_{mid}$ is given by placing $h$ in equation \ref{eq:tau_final} as follows:
\begin{equation} \label{eq:tau_mid_final}
    \tau_{mid} = \frac{4r^3 h}{3 w^2 D}
\end{equation}

In the $tail$ regime, the limiting factors are both the pore height $(h)$ and the sensor distance $(d)$. We also know that the characteristic length in 2D diffusion is $\sqrt{2}$ times the characteristic length in 1D diffusion. Hence, the characteristic length in the $tail$ regime will be given by $l_{char} = h + \sqrt{2}d$. Substituting $l_{char}$ for $h$ in equation \ref{eq:tau_final} gives us the following expression for $\tau_{tail}$

\begin{equation} \label{eq:tau_tail_final}
    \tau_{tail} = \frac{4r^3(h+\sqrt{2}d)}{3 w^2 D}
\end{equation}

\begin{figure*}[ht]
\includegraphics[width=\textwidth]{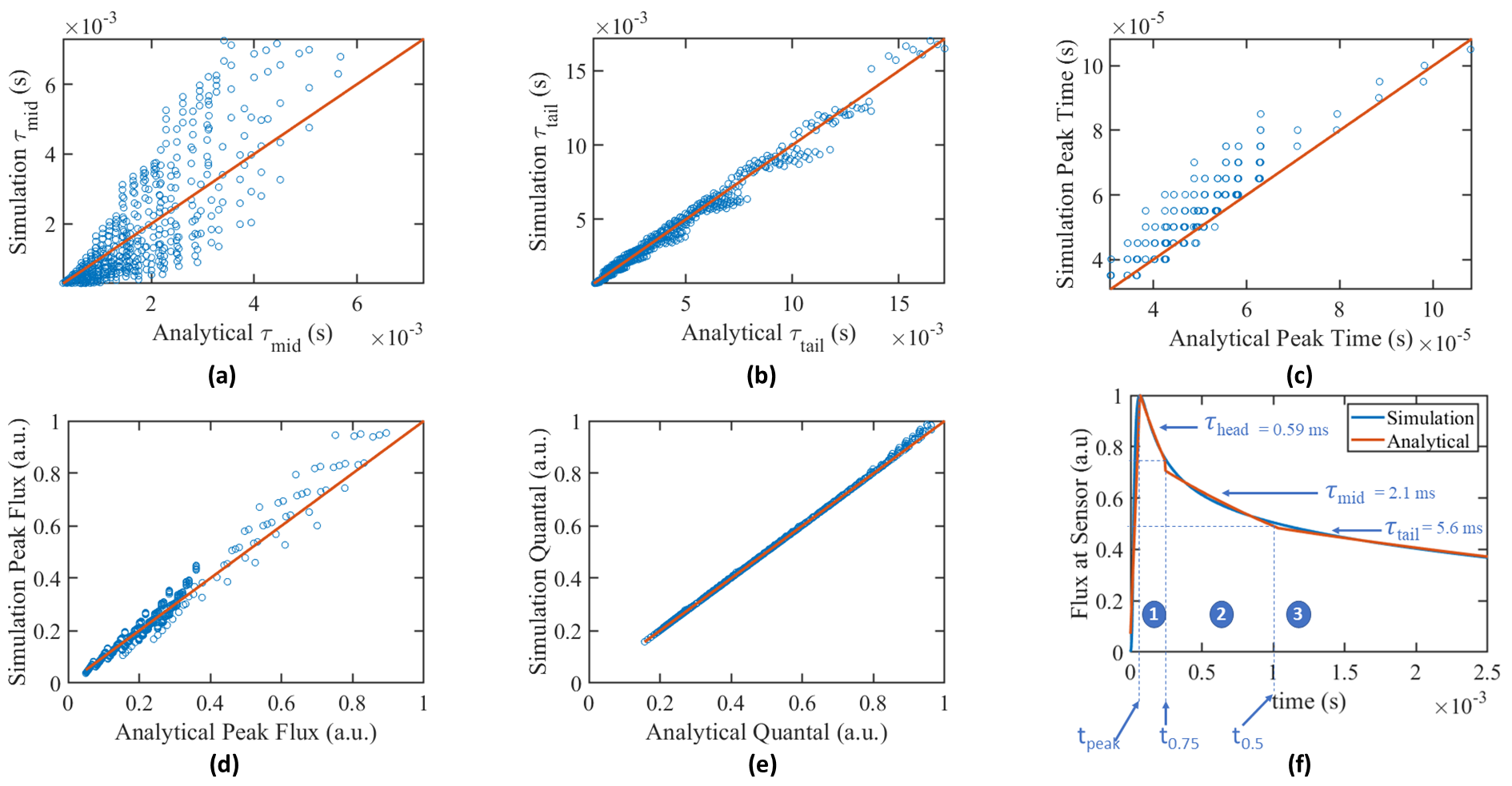}
\caption{\label{fig:scatter_plots} Scatter plots comparing  Output Parameter values as predicted from the analytical model with the values obtained from the simulation (a) $\tau_{mid}$ - Time Constant from the mid region (equation \ref{eq:tau_mid_final}) (b) $\tau_{tail}$ - Time constant in the tail region, (equation \ref{eq:tau_tail_final}) (c) $t_{peak}$ - Time for Flux to reach peak (equation \ref{eq:peak_time_final}) (d) $F_{peak}$ - Peak Flux at the sensor (equation \ref{eq:peak_flux_eq}) (e) $Q$ - Quantal or total number of neurotransmitters received (f) Simulation flux vs Analytical flux based on three time regimes and three time constants for the following parameters - $r = 100 nm, w = 40 nm, h = 10 nm,$ and $d = 10 nm$
}
\end{figure*}

\textbf{B. Peak time} Apart from the time constants, the time at which the electrode current reaches a maximum is another parameter which can be unambiguously obtained from experiments. As discussed before, at the instant of pore opening, there is a rush of particles from the vesicle into the cylindrical pore and then outside the neuron towards the sensor. Here there are two distinct regions of diffusion,namely the pore and the gap between the neuron and sensor. Both these regions are characterised by different characteristic lengths given by $l_{pore} = h$, and $l_{gap} = \sqrt{2}d$. In the previous section, we saw that the diffusion is limited by $l_{pore}$ in the $mid$ regime and  by the sum of both in the $tail$ regime. However, in the $head$ regime, the characteristic length is a combination of the two in such a way that at $t=0$, the characteristic length $(l_{char})$ is equivalent to $l_{pore}$ and as $t \rightarrow \infty$, the characteristic length tends to $l_{pore} + l_{gap}$. Based on these boundary conditions, we can define an expression for characteristic length as a function of time as

\begin{equation} \label{eq:char_length}
    l_{char}(t) = h+(1-e^{-\frac{t}{\tau}})\sqrt{2}d
\end{equation}

Now to arrive at an expression for the peak time $(t_{peak})$, we make the assumption that from $t=0$ to $t=t_{peak}$, it is 1D diffusion from a constant source at the pore and vesicle interface. Hence, the peak will happen at a time which is the solution to the following equation 
\begin{equation} \label{eq:peak_time_time}
    l_{char}(t)^2 = 2Dt
\end{equation}

Expanding the above equation, using the Taylor series expansion for $e^{-x}$ till the linear term (since $\frac{t}{\tau} <<< 1$), and re-arranging the terms we get the following expression for $t_{peak}$
\begin{equation} \label{eq:peak_time_final}
    t_{peak} =\tau \frac{h^2 + 2d^2 + \sqrt{2}dh}{2D\tau + 2d^2 - \sqrt{2}dh}
\end{equation}

\textbf{C. Peak Flux} Based on the discussion above, and known solutions of 1D diffusion from a constant source, we can estimate the peak flux at the exit of the pore at the cell membrane. We can also see that the open 2D diffusion between the pore and the sensor will reduce the peak flux. However, we can assume that the peak flux at the sensor will be a fraction of the peak flux at the exit of the pore. Hence, we can estimate that $F_{peak}(sensor) = c_2 F_{peak}(pore)$, where $c_2$ is a constant less than one. Based on the expression for peak flux from a constant point source, and multiplying it by the area of the exit surface of the pore, we get the following expression
\begin{equation} \label{eq:peak_flux_eq}
F_{peak}(sensor) = c_2 e^{-1/2}\sqrt{2\pi}\frac{C_sDw^2}{h+\sqrt{2}d}
\end{equation}
where $w$ is the width of the pore, $h$ is the height of the pore, $d$ is the sensor distance, $C_s$ is the concentration at source, $D$ is the diffusion constant, and $c_{2}$ is a fraction to be determined.

\section{Simulation Results}

Transient diffusion equation was numerically solved for the model system with cylindrical symmetry (see eq. \ref{eq:diff_eq2}). Finite difference scheme with BDF2 time integration was employed for numerical solution of the discretized equations. Extensive simulations were performed to explore the influence of geometry by varying the following four parameters: (i) Vesicle Radius $(r)$ from $60$ to $110 nm$ in steps of $5 nm$, (ii) Pore Width $(w)$ from $15$ to $75 nm$ in steps of $5 nm$, (iii) Pore Height $(h)$ from $7$ to $10 nm$ in steps of $1 nm$, and (iv) Sensor Distance $(d)$ from $7$ to $15 nm$ in steps of $1 nm$. We kept the Sensor Length $(l)$ fixed at $200 nm$ to model the case of sensor being much larger than the pore width, pore height and sensor distance, i.e. $l >> w,h,$ and $d$, and the assumed diffusion constant $(D)$ is $3.5 \times 10^{-8} cm^2/s$. We ran the simulations for a time duration of $5 ms$.  

Based on the simulations, we computed five output parameters for each combination of the input parameters, excluding scenarios where pore width is less than $15\%$ of the vesicle radius, i.e. $w < 0.15r$. We then compared the values of the parameters from the simulations with the predicted values from the analytical expressions derived in the previous section. From the scatter plots in Fig. \ref{fig:scatter_plots}, we find that the expressions derived in the previous section compares well with the simulated values for all the five output parameters shown.  Further, the results also indicate that there are clearly three distinct time constants in the signal as predicted by the analytical model. This feature is evident in Fig. \ref{fig:scatter_plots}f with the three time constants as $\tau_{head} = 0.59 ms$, $\tau_{mid} = 2.1 ms$, and $\tau_{tail} = 5.6 ms$.

\section{Conclusions}
In this manuscript, we  developed predictive analytical models to anticipate the key features of exocytosis process. Our model predictions on multi-exponential decay, time constants, peak flux, quantal, etc. are well supported by results from detailed numerical simulations. In future, the model could be updated for scenarios with dynamic pore. As such, this work could enable facile analysis of experimental data and back extract important information regarding the key processes involved in exocytosis.

\section{Acknowledgements}
We would like to acknowledge and thank Prof. Bhaskaran Muralidharan of the Electrical Engineering Department, IIT Bombay for insightful discussions.

\section{References}
\bibliography{predictivemodel}

\end{document}